
\magnification \magstep1
\raggedbottom
\openup 2\jot
\voffset6truemm
\headline={\ifnum\pageno=1\hfill\else
\hfill {\it Non-Local Properties in Euclidean Quantum Gravity}
\hfill \fi}
\def\cstok#1{\leavevmode\thinspace\hbox{\vrule\vtop{\vbox{\hrule\kern1pt
\hbox{\vphantom{\tt/}\thinspace{\tt#1}\thinspace}}
\kern1pt\hrule}\vrule}\thinspace}
\centerline {\bf NON-LOCAL PROPERTIES IN}
\centerline {\bf EUCLIDEAN QUANTUM GRAVITY}
\vskip 1cm
\centerline {GIAMPIERO ESPOSITO}
\vskip 1cm
\centerline {\it Istituto Nazionale di Fisica Nucleare,
Sezione di Napoli}
\centerline {\it Mostra d'Oltremare Padiglione 20,
80125 Napoli, Italy}
\vskip 0.3cm
\centerline {\it Dipartimento di Scienze Fisiche}
\centerline {\it Mostra d'Oltremare Padiglione 19,
80125 Napoli, Italy}
\vskip 1cm
\noindent
{\bf Abstract.} In the one-loop approximation for Euclidean
quantum gravity, the boundary conditions which are completely
invariant under gauge transformations of metric perturbations
involve both normal and tangential derivatives of the metric
perturbations $h_{00}$ and $h_{0i}$,
while the $h_{ij}$ perturbations and the
whole ghost one-form are set to zero at the boundary.
The corresponding
one-loop divergency for pure gravity has been recently
evaluated by means of analytic techniques. It now remains to
compute the contribution of all perturbative modes of gauge
fields and gravitation to the one-loop effective action for
problems with boundaries. The functional determinant has a
non-local nature, independently of boundary conditions.
Moreover, the analysis of one-loop divergences for
supergravity with non-local boundary conditions has not
yet been completed and is still under active investigation.
\vskip 1cm
\noindent
To appear in Proceedings of the {\it Third Workshop on Quantum
Field Theory under the Influence of External Conditions},
Leipzig, September 1995 (DSF preprint 95/37, GR-QC 9508056).
\vskip 100cm
\leftline {\bf 1. Introduction}
\vskip 1cm
\noindent
This paper is devoted to local and non-local properties which are
relevant for the analysis of Euclidean quantum gravity in the
presence of boundaries. Before presenting the technical details,
it is necessary to describe why boundaries are so important in
quantum gravity. As far as we can see, there are at least two
main motivations:
\vskip 0.3cm
\noindent
(i) The propagator of quantum gravity may be expressed formally
as a path integral over all Riemannian four-geometries matching
the boundary data on two (compact) Riemannian three-geometries
$\Bigr(\Sigma_{1},h_{1}\Bigr)$ and $\Bigr(\Sigma_{2},h_{2}\Bigr)$,
where $h_{i}$ is the metric induced on the surface $\Sigma_{i}$,
with $i=1,2$. It is then necessary to understand how to fix the
boundary data on $\Bigr(\Sigma_{1},h_{1}\Bigr)$ and
$\Bigr(\Sigma_{2},h_{2}\Bigr)$. In quantum cosmology, this
analysis leads to a prescription for the quantum state of the
universe, i.e. a functional of the three-geometry which solves
the Wheeler-DeWitt equation and represents the probability
amplitude of having data on a compact Riemannian three-geometry
$(\Sigma,h)$ [1,2]. These data consist of the metric
configuration on $(\Sigma,h)$, and of matter field configurations
(e.g. fermionic fields or bosonic gauge fields).
\vskip 0.3cm
\noindent
(ii) The effective action remains the main tool of perturbative
quantum field theory [3-5]. Its one-loop approximation contains
relevant information about trace anomalies
and one-loop divergences [6,7], and is at the
heart of symmetry-breaking phenomena [8-13]. The general form of
volume terms in the corresponding asymptotic heat kernel
has been obtained, after many years of dedicated work, by DeWitt,
Gilkey, Avramidi [14-16]. In the presence of boundaries, surface
terms occur which have a rich geometric structure and are
necessary to obtain the correct values of the trace anomalies
and to investigate the non-local nature of the one-loop effective
action. For spinor fields, gauge fields and gravitation, the
correct values of these one-loop divergences have been obtained for
the first time only very recently, by using analytic or geometric
techniques [6,17-25]. There is agreement, by now, between the
analytic (mode-by-mode) and geometric (space-time covariant)
calculations of the trace anomalies for gauge fields
and one-loop divergences for
gravitation subject to local boundary conditions, when the
Faddeev-Popov formalism is used with manifestly covariant
gauges [22-25]. However,
the presence of boundaries leads to severe
technical complications, and the geometric form of such divergences
with non-covariant gauges and other families of boundary
conditions is not yet completely understood.

In section 2 we derive in detail a set of mixed boundary
conditions in Euclidean quantum gravity. In section 3 we
discuss the open problems in this branch of perturbative
quantum gravity.
\vskip 1cm
\leftline {\bf 2. Mixed Boundary Conditions for Euclidean
Quantum Gravity}
\vskip 1cm
\noindent
For gauge fields and gravitation, the boundary conditions
are mixed, in that some components of the field (more
precisely, a one-form or a two-form) obey a set of boundary
conditions, and the remaining part of the field obeys another
set of boundary conditions. Moreover, the boundary conditions
are invariant under local gauge transformations providing
suitable boundary conditions are imposed on the corresponding
ghost zero-form or one-form.

We are here interested in the derivation of mixed boundary
conditions for Euclidean quantum gravity. The knowledge of
the classical variational problem, and the principle of
gauge invariance, are enough to lead to a highly non-trivial
quantum boundary-value problem. Indeed, it is by now well known
that, if one fixes the three-metric at the boundary in
general relativity, the corresponding variational problem
is well posed and leads to the Einstein equations, providing
the Einstein-Hilbert action is supplemented by a boundary
term whose integrand is proportional to the trace of the
second fundamental form [6,26-27]. In the corresponding
quantum boundary-value problem, which is relevant for the
one-loop approximation in quantum gravity [6], the perturbations
$h_{ij}$ of the induced three-metric are set to zero at the
boundary. Moreover, the whole set of metric perturbations
$h_{\mu \nu}$ are subject to the so-called {\it gauge}
transformations [25]
$$
{\widehat h}_{\mu \nu} \equiv h_{\mu \nu}
+\nabla_{(\mu} \; \varphi_{\nu)}
\; \; \; \; ,
\eqno (2.1)
$$
where $\nabla$ is the Levi-Civita connection of the background
four-geometry with metric $g$, and $\varphi_{\nu}dx^{\nu}$
is the ghost one-form [25]. In geometric language, the
difference between ${\widehat h}_{\mu \nu}$ and
$h_{\mu \nu}$ is given by the Lie derivative along $\varphi$
of the four-metric $g$.

For problems with boundaries, Eq. (2.1) implies that
$$
{\widehat h}_{ij}=h_{ij}+\varphi_{(i \mid j)}
+K_{ij} \varphi_{0}
\; \; \; \; ,
\eqno (2.2)
$$
where the stroke denotes, as usual, three-dimensional
covariant differentiation tangentially with respect to the
intrinsic Levi-Civita connection of the boundary, while
$K_{ij}$ is the extrinsic-curvature tensor of the boundary.
Of course, $\varphi_{0}$ and $\varphi_{i}$ are the normal
and tangential components of the ghost one-form, respectively.
Note that boundaries make it necessary to perform a 3+1
split of space-time geometry and physical fields.
As such, they introduce non-covariant elements in the analysis
of problems relevant for quantum gravity. This seems to be
an unavoidable feature, although the boundary conditions may
be written in a covariant way [28].

In the light of (2.2), the boundary conditions
$$
\Bigr[h_{ij}\Bigr]_{\partial M}=0
\eqno (2.3a)
$$
are gauge invariant, i.e.
$$
\Bigr[{\widehat h}_{ij}\Bigr]_{\partial M}=0
\; \; \; \; ,
\eqno (2.3b)
$$
if and only if the whole ghost one-form obeys homogeneous
Dirichlet conditions, so that
$$
\Bigr[\varphi_{0}\Bigr]_{\partial M}=0
\; \; \; \; ,
\eqno (2.4)
$$
$$
\Bigr[\varphi_{i}\Bigr]_{\partial M}=0
\; \; \; \; .
\eqno (2.5)
$$
To prove necessity and sufficiency of the conditions (2.4)-(2.5),
one has to bear in mind the independent expansions in harmonics
of $\varphi_{0}$ and $\varphi_{i}$. These obey a factorization
property, and hence the three-dimensional covariant derivatives
only act on the spatial harmonics, so that $\varphi_{(i \mid j)}$
vanishes at the boundary if and only if (2.5) holds [25].

The problem now arises to impose boundary conditions on the
remaining set of metric perturbations. The key point is to
make sure that the invariance of such boundary conditions under
the transformations (2.1) is again guaranteed by (2.4)-(2.5),
since otherwise one would obtain incompatible sets of
boundary conditions on the ghost one-form. Indeed, on using
the Faddeev-Popov formalism for the amplitudes of quantum
gravity, it is necessary to use a gauge-averaging term in
the Euclidean action, of the form [24]
$$
I_{\rm g.a.} \equiv {1\over 32 \pi G \alpha}
\int_{M}\Phi_{\nu}\Phi^{\nu}\sqrt{{\rm det} \; g} \; d^{4}x
\; \; \; \; ,
\eqno (2.6)
$$
where $\Phi_{\nu}$ is any relativistic gauge-averaging
functional which leads to self-adjoint elliptic operators
on metric and ghost perturbations. In particular, if the
de Donder gauge is chosen [24],
$$
\Phi_{\nu}^{dD}(h) \equiv \nabla^{\mu} \Bigr(h_{\mu \nu}
-{1\over 2}g_{\mu \nu} g^{\rho \sigma}h_{\rho \sigma}\Bigr)
\; \; \; \; ,
\eqno (2.7)
$$
one finds that [25]
$$
\Phi_{\nu}^{dD}(h)-\Phi_{\nu}^{dD}(\widehat h)
=-{1\over 2}\Bigr(g_{\mu \nu}\cstok{\ }+R_{\mu \nu}\Bigr)
\varphi^{\mu}={\lambda \over 2}\varphi_{\nu}
\; \; \; \; ,
\eqno (2.8)
$$
where $\cstok{\ } \equiv g^{\mu \nu}\nabla_{\mu}\nabla_{\nu}$,
$R_{\mu \nu}$ is the Ricci tensor of the background, and
$\lambda$ denotes the eigenvalues of the elliptic operator
$-\Bigr(g_{\mu \nu}\cstok{\ }+R_{\mu \nu}\Bigr)$.
Indeed, our notation in the second equality of (2.8) is loose,
and it is enough to emphasize the elliptic nature of the
operator acting on the ghost one-form. Thus, if one
imposes the boundary conditions
$$
\Bigr[\Phi_{0}^{dD}(h)\Bigr]_{\partial M}=0
\; \; \; \; ,
\eqno (2.9a)
$$
$$
\Bigr[\Phi_{i}^{dD}(h)\Bigr]_{\partial M}=0
\; \; \; \; ,
\eqno (2.10a)
$$
their invariance under (2.1) is guaranteed when (2.4)-(2.5)
hold, by virtue of (2.8). Hence one also has
$$
\Bigr[\Phi_{0}^{dD}(\widehat h)\Bigr]_{\partial M}=0
\; \; \; \; ,
\eqno (2.9b)
$$
$$
\Bigr[\Phi_{i}^{dD}(\widehat h)\Bigr]_{\partial M}=0
\; \; \; \; .
\eqno (2.10b)
$$
Note that the boundary conditions on the ghost one-form
become redundant if one also imposes the conditions (2.3b),
(2.9b) and (2.10b). Nevertheless, we shall always write them
explicitly, since the ghost one-form plays a key role
in quantum gravity.

The most general scheme does {\it not}
depend on the choice of the de Donder term, so that it
relies on (2.3a)-(2.3b), (2.4)-(2.5), jointly with [7,25]
$$
\Bigr[\Phi_{0}(h)\Bigr]_{\partial M}=0
\; \; \; \; ,
\eqno (2.11a)
$$
$$
\Bigr[\Phi_{0}(\widehat h)\Bigr]_{\partial M}=0
\; \; \; \; ,
\eqno (2.11b)
$$
$$
\Bigr[\Phi_{i}(h)\Bigr]_{\partial M}=0
\; \; \; \; ,
\eqno (2.12a)
$$
$$
\Bigr[\Phi_{i}(\widehat h)\Bigr]_{\partial M}=0
\; \; \; \; .
\eqno (2.12b)
$$
Again, it is enough to write (2.3a), (2.11a), (2.12a),
(2.4)-(2.5), or (2.3a)-(2.3b) jointly with (2.11a)-(2.11b)
and (2.12a)-(2.12b).

In the particular and relevant case of flat Euclidean four-space
bounded by a three-sphere [6,29], the de Donder gauge and the
boundary conditions (2.3a), (2.9a), (2.10a) lead to [25]
$$
\left[{\partial h_{00}\over \partial \tau}
+{6\over \tau}h_{00}-{\partial \over \partial \tau}
\Bigr(g^{ij}h_{ij}\Bigr)+{2\over \tau^{2}}
h_{0i}^{\; \; \; \mid i} \right]_{\partial M}=0
\; \; \; \; ,
\eqno (2.13)
$$
$$
\left[{\partial h_{0i}\over \partial \tau}
+{3\over \tau}h_{0i}-{1\over 2}{\partial h_{00}\over
\partial x^{i}}\right]_{\partial M}=0
\; \; \; \; ,
\eqno (2.14)
$$
where $\tau$ is the Euclidean-time coordinate, which here
represents the radius of three-spheres centred at the origin.
These boundary conditions have three basic properties:
\vskip 0.3cm
\noindent
(i) They involve both normal and tangential derivatives of
the $h_{00}$ and $h_{0i}$ metric perturbations.
\vskip 0.3cm
\noindent
(ii) The ghost boundary conditions cannot be
expressed in terms of complementary projection operators
on $\varphi_{0}$ and $\varphi_{i}$, and are instead Dirichlet
on both $\varphi_{0}$ and $\varphi_{i}$.
\vskip 0.3cm
\noindent
(iii) They can be combined with non-local boundary conditions
in $N=1$ supergravity, when half of the tangential part of
the gravitino potential is set to zero at the boundary.
This is a non-local operation, since it makes it necessary
to pick out the modes which multiply harmonics having
positive eigenvalues for the intrinsic three-dimensional
Dirac operator at the boundary. What is non-local is the
separation of the spectrum of an elliptic operator into
its positive and negative parts, and this leads to one
of the two possible sets of mixed boundary conditions
for spin-3/2 potentials [6,18].

When a three-sphere bounds flat Euclidean four-space, the
symmetries of the problem and the use of zeta-function
regularization [30,31] make it possible to evaluate the
corresponding one-loop divergency
for pure gravity, which is found to be [25]
$$
\zeta(0)=-{241\over 90}
\; \; \; \; .
\eqno (2.15)
$$
\vskip 1cm
\leftline {\bf 3. Achievements and Unsolved Problems}
\vskip 1cm
\noindent
Over the past six years, impressive progress has been made
in our understanding of boundary terms in the asymptotic
heat kernel. Thus, the trace anomalies for massless scalar fields
subject to Dirichlet, Neumann or Robin boundary conditions,
or massless spin-1/2 fields with local or spectral boundary conditions,
or Euclidean Maxwell theory in vacuum with magnetic or electric
boundary conditions, or one-loop divergences for
linearized gravity with three different
sets of mixed boundary conditions, are by now well known
[17-25,32,33]. Moreover, when boundaries are present, the
$\zeta'(0)$ values for scalar and spin-1/2 fields, and the
contributions of physical degrees of freedom to
$\zeta'(0)$ for spin-1, spin-3/2 and spin-2 fields have also
been obtained [34-37]. Despite this very encouraging progress,
where one should also mention the contribution of physical
degrees of freedom to the one-loop divergency for gravitino
potentials [6,38], many important problems remain unsolved.
They are as follows.
\vskip 0.3cm
\noindent
(i) What is the {\it geometric} form of the one-loop divergency
for linearized gravity subject to the boundary conditions
of section 2, and of the trace anomaly for
massless spin-1/2 fields subject
to non-local boundary conditions ? Can one find a suitable
generalization of the Schwinger-DeWitt ansatz (cf. [39,40]) ?
\vskip 0.3cm
\noindent
(ii) What is the contribution of gauge modes and ghost modes
to the one-loop divergency for gravitino potentials with non-local
boundary conditions ?
\vskip 0.3cm
\noindent
(iii) What is the correct form of $\zeta'(0)$ for gauge fields
and gravitation, when {\it all} perturbative modes are taken
into account in the presence of boundaries ? [The work in
[37] is restricted to the so-called physical degrees of
freedom, e.g. the transverse part of the electromagnetic
potential, or transverse-traceless perturbations for
pure gravity.]
\vskip 0.3cm
\noindent
(iv) Can one evaluate the one-loop divergences with {\it arbitrary}
relativistic gauges ? [These gauges [41,42] lead to non-minimal
operators, and the presence of boundaries makes it very
difficult to perform a mode-by-mode analysis of the quantized
field in such a case.]
\vskip 0.3cm
\noindent
(v) Can one prove essential self-adjointness of the various
elliptic operators occurring in the semi-classical
amplitudes ?
\vskip 0.3cm
\noindent
(vi) Can one pick out a preferred choice of mixed boundary
conditions for Euclidean quantum gravity, among the four
different sets studied so far in the literature
[24,25,33,43-46] ?

To make sense of the quantum state of the universe [2],
of the path-integral approach to quantum gravity [47] and
of the corresponding semi-classical approximation [6],
it is essential to understand the various aspects of the
problem of boundary conditions, i.e. the form of the
boundary, the four-geometries summed over in the path
integral and the boundary data chosen in the one-loop
calculation. For this purpose, the old geometrodynamical
framework [1], jointly with the key principles of modern
quantum field theory, e.g. gauge invariance and BRST
symmetry, is still the most appropriate for the active
investigation of these issues. Hence we hope that, in
the near future, the scientific community will come to
appreciate how deep are the issues raised, and possibly
solved, by a rigorous analysis of quantum field theories
in the presence of boundaries [6,28].
\vskip 0.3cm
\leftline {\bf Acknowledgments}
\vskip 0.3cm
\noindent
I would like to thank Michael Bordag and Klaus Kirsten
for their warm hospitality during the Leipzig Workshop.
I am indebted to Peter D'Eath, Alexander Kamenshchik,
Igor Mishakov and Giuseppe Pollifrone for scientific
collaboration on the evaluation of one-loop divergences,
and to Andrei Barvinsky, Stuart Dowker, Hugh Luckock,
Ian Moss, Stephen Poletti and Dmitri Vassilevich
for correspondence. I am also indebted to Ivan Avramidi
for comments and criticisms which led to an
improvement of the original manuscript.
The work of Ian Moss [48,49] has been
of invaluable importance to obtain a correct approach
to one-loop quantum cosmology.
Last, but not least, I am grateful to
the European Union for partial support under the Human
Capital and Mobility Programme.
\vskip 0.3cm
\leftline {\bf References}
\vskip 0.3cm
\item {[1]}
Wheeler J. A. (1962) {\it Geometrodynamics}
(New York: Academic Press).
\item {[2]}
Hawking S. W. (1984) {\it Nucl. Phys.} {\bf B 239}, 257.
\item {[3]}
Vilkovisky G. A. (1984) {\it Nucl. Phys.} {\bf B 234}, 125.
\item {[4]}
DeWitt B. S. (1987) {\it The Effective Action}, in
{\it Architecture of Fundamental Interactions at Short
Distances}, Les Houches Session XLIV, eds. P. Ramond and
R. Stora (Amsterdam: North-Holland) p. 1023.
\item {[5]}
Buchbinder I. L., Odintsov S. D. and Shapiro I. L. (1992)
{\it Effective Action in Quantum Gravity} (Bristol: IOP).
\item {[6]}
Esposito G. (1994) {\it Quantum Gravity, Quantum Cosmology
and Lorentzian Geometries}, Lecture Notes in Physics,
New Series m: Monographs, Vol. m12, second corrected and
enlarged edition (Berlin: Springer-Verlag).
\item {[7]}
Esposito G. {\it The Impact of Quantum Cosmology on Quantum
Field Theory}, to appear in Proceedings of the 1995 Moscow
Quantum Gravity Seminar (DSF preprint 95/27, GR-QC 9506046).
\item {[8]}
Coleman S. and Weinberg E. (1973) {\it Phys. Rev.}
{\bf D 7}, 1888.
\item {[9]}
Allen B. (1983) {\it Nucl. Phys.} {\bf B 226}, 228.
\item {[10]}
Buccella F., Esposito G. and Miele G. (1992) {\it Class.
Quantum Grav.} {\bf 9}, 1499.
\item {[11]}
Esposito G., Miele G. and Rosa L. (1993) {\it Class.
Quantum Grav.} {\bf 10}, 1285.
\item {[12]}
Esposito G., Miele G. and Rosa L. (1994) {\it Class.
Quantum Grav.} {\bf 11}, 2031.
\item {[13]}
Esposito G., Miele G., Rosa L. and Santorelli P.
{\it Quantum Effects in FRW Cosmologies}
(DSF preprint 95/6, GR-QC 9508010).
\item {[14]}
DeWitt B. S. (1965) {\it Dynamical Theory of Groups and
Fields} (New York: Gordon and Breach).
\item {[15]}
Gilkey P. B. (1975) {\it J. Diff. Geom.} {\bf 10}, 601.
\item {[16]}
Avramidi I. (1991) {\it Nucl. Phys.} {\bf B 355}, 712.
\item {[17]}
D'Eath P. D. and Esposito G. (1991) {\it Phys. Rev.}
{\bf D 43}, 3234.
\item {[18]}
D'Eath P. D. and Esposito G. (1991) {\it Phys. Rev.}
{\bf D 44}, 1713.
\item {[19]}
Kamenshchik A. Yu. and Mishakov I. V. (1992)
{\it Int. J. Mod. Phys.} {\bf A 7}, 3713.
\item {[20]}
Kamenshchik A. Yu. and Mishakov I. V. (1993)
{\it Phys. Rev.} {\bf D 47}, 1380.
\item {[21]}
Kamenshchik A. Yu. and Mishakov I. V. (1994)
{\it Phys. Rev.} {\bf D 49}, 816.
\item {[22]}
Moss I. G. and Poletti S. J. (1994) {\it Phys. Lett.}
{\bf B 333}, 326.
\item {[23]}
Esposito G., Kamenshchik A. Yu., Mishakov I. V. and
Pollifrone G. (1994) {\it Class. Quantum Grav.} {\bf 11}, 2939.
\item {[24]}
Esposito G., Kamenshchik A. Yu., Mishakov I. V. and
Pollifrone G. (1994) {\it Phys. Rev.} {\bf D 50}, 6329.
\item {[25]}
Esposito G., Kamenshchik A. Yu., Mishakov I. V. and
Pollifrone G. (1995) {\it Phys. Rev.} {\bf D 52}, 3457.
\item {[26]}
York J. W. (1972) {\it Phys. Rev. Lett.} {\bf 28}, 1072.
\item {[27]}
York J. W. (1986) {\it Found. Phys.} {\bf 16}, 249.
\item {[28]}
Esposito G., Kamenshchik A. Yu. and Pollifrone G.
{\it Quantum Field Theories in the Presence of Boundaries}
(book in preparation).
\item {[29]}
Schleich K. (1985) {\it Phys. Rev.} {\bf D 32}, 1889.
\item {[30]}
Hawking S. W. (1977) {\it Commun. Math. Phys.}
{\bf 55}, 133.
\item {[31]}
Barvinsky A. O., Kamenshchik A. Yu. and Karmazin I. P.
(1992) {\it Ann. Phys.} {\bf 219}, 201.
\item {[32]}
Bordag M., Elizalde E. and Kirsten K. {\it Heat-Kernel
Coefficients of the Laplace Operator on the D-Dimensional
Ball} (UB-ECM-PF preprint 95/3, HEP-TH 9503023).
\item {[33]}
Esposito G. and Kamenshchik A. Yu. {\it Mixed Boundary
Conditions in Euclidean Quantum Gravity} (DSF preprint
95/23, GR-QC 9506092).
\item {[34]}
Dowker J. S. and Apps J. S. (1995) {\it Class. Quantum
Grav.} {\bf 12}, 1363.
\item {[35]}
Bordag M., Geyer B., Kirsten K. and Elizalde E.
{\it Zeta-Function Determinant of the Laplace Operator
on the D-Dimensional Ball}
(UB-ECM-PF preprint 95/10, HEP-TH 9505157).
\item {[36]}
Dowker J. S. {\it Spin on the 4-Ball} (MUTP preprint 95/13,
HEP-TH 9508082).
\item {[37]}
Kirsten K. and Cognola G. {\it Heat-Kernel
Coefficients and Functional Determinants for Higher-Spin
Fields on the Ball} (UTF preprint 354, HEP-TH 9508088).
\item {[38]}
Esposito G. (1994) {\it Int. J. Mod. Phys.} {\bf D 3}, 593.
\item {[39]}
McAvity D. M. and Osborne H. (1991) {\it Class. Quantum
Grav.} {\bf 8}, 603.
\item {[40]}
McAvity D. M. and Osborne H. (1991) {\it Class. Quantum
Grav.} {\bf 8}, 1445.
\item {[41]}
Esposito G. (1994) {\it Class. Quantum Grav.} {\bf 11}, 905.
\item {[42]}
Esposito G., Kamenshchik A. Yu., Mishakov I. V. and
Pollifrone G. (1995) {\it Phys. Rev.} {\bf D 52}, 2183.
\item {[43]}
Barvinsky A. O. (1987) {\it Phys. Lett.} {\bf B 195}, 344.
\item {[44]}
Luckock H. C. (1991) {\it J. Math. Phys.} {\bf 32}, 1755.
\item {[45]}
Moss I. G. and Poletti S. J. (1990) {\it Nucl. Phys.}
{\bf B 341}, 155.
\item {[46]}
Marachevsky V. N. and Vassilevich D. V. (1995)
{\it Diffeomorphism Invariant Eigenvalue Problem for Metric
Perturbations in a Bounded Region} (paper in preparation).
\item {[47]}
Hawking S. W. (1979) {\it The Path-Integral Approach to
Quantum Gravity}, in {\it General Relativity, an Einstein
Centenary Survey}, eds. W. Israel and S. W. Hawking
(Cambridge: Cambridge University Press) p. 746.
\item {[48]}
Moss I. G. (1989) {\it Class. Quantum Grav.} {\bf 6}, 759.
\item {[49]}
Moss I. G. (1995) {\it One-Loop Quantum Cosmology and the
Vanishing Ghost}, talk given at the 1995
Moscow Quantum Gravity Seminar.

\bye